\documentclass[12pt]{article} 
\usepackage{amssymb,latexsym} 
\usepackage{amsmath} 

\columnwidth\textwidth

\begin{document}

\newtheorem{df}{Definition} 
\newtheorem{thm}{Theorem} 
\newtheorem{lem}{Lemma} 

\begin{titlepage} 

\noindent 

\vspace*{1cm} 
\begin{center} 
{\LARGE Freedom in Nature} 

\vspace{2cm} 

P. H\'{a}j\'{\i}\v{c}ek \\ 
Institute for Theoretical Physics \\ 
University of Bern \\ 
Sidlerstrasse 5, CH-3012 Bern, Switzerland \\ 
tel.\ 0041318690727 \\ 
fax 0041316313821 \\ 
e-mail hajicek@itp.unibe.ch \\ 
\vspace{0.5cm}

October 2008 \\  
\vspace*{1cm} 
\nopagebreak[4] 

\end{center} 

\end{titlepage} 

\noindent {\bf Abstract:}
The paper starts with the proposal that the cause of the apparent insolubility
of the free-will problem are several popular but strongly metaphysical notions
and hypotheses. To reduce the metaphysics, some ideas are borrowed from
physics. A concept of event causality is discussed. The importance of Hume's
Principle of Causality is stressed and his Principle of Causation is
weakened. The key concept of the paper, the so-called relative freedom, is
also suggested by physics. It is a kind of freedom that can be observed
everywhere in nature. Turning to biology, incomplete knowledge is defined for
all organisms. They cope with the problem by Popper's trial and error
processes. One source of their success is the relative freedom of choice from
the basic option ranges: mutations, motions and neural connections. Finally,
the conjecture is adopted that communicability can be used as a criterion of
consciousness and free will is defined as a conscious version of relative
freedom. The resulting notion is logically self-consistent and it describes an
observable phenomenon that agrees with our experience. 
\\

\noindent {\bf Key words:} physics; causality; free will; memory;
consciousness; incomplete knowledge; trial and error 

\newpage 

\section{Introduction} 
If anything happens, we have the tendency to ask two questions: 'What is the
cause of it?' and 'What am I to do?' The questions originate in some deep
(perhaps unconscious) convictions. The first is that every event must have a
cause and the second that we can influence things by a suitably chosen
action. In our practical life, the resulting two activities go hand in
hand. For example, the knowledge of causes and effects improves our ability to
choose actions so that they have the results we desire. 

However, if we try to build a serious theory that would provide a more exact
formulation of the convictions as well as underpin and explain them, we run
into well-known difficulties. 

To formulate the first conviction, most thinkers arrive at the principle
(Platon, Timaios): 
\begin{quote}
Everything what happens must happen by a cause because it is impossible that
anything comes into being without cause. 
\end{quote}
A more recent formulation is that by Laplace (1820):
\begin{quote}
An intellect which at any given moment knew all the forces that animate Nature
and the mutual positions of the beings that comprise it, if this intellect
were wast enough to submit its data to analysis, could condense into a single
formula the movement of the greatest bodies of the universe and that of the
lightest atom: for such intellect nothing would be uncertain; and the future
just like the past would be present before its eyes. 
\end{quote}
This principle is called {\em determinism}; it gives a more exact formulation
of the first conviction. The antithesis of determinism is that there are
events that do not have causes. We shall call such events {\em random}. The
antithesis has become popular in physics by the influence of quantum
mechanics. A question is, how it can be made compatible with our tendency to
look for causes. 

A more detailed formulation of the second conviction and of the problem to
understand it is the following (Double 1991): 
\begin{quote}
Any acceptable explication of free will:
\begin{enumerate}
\item must entail that free person could have chosen otherwise,
\item must explicate the control that free will requires,
\item must explicate the "sensibleness" or "rationality" that free will
  involves. 
\end{enumerate}.
\end{quote}

The first point assumes that there is a range of options from which one
chooses. An important question is which conditions determine the range. An
opinion that is often met can be characterized by adding the words '{\em all}
other conditions remaining the same' (Searle (1984), P. 95) to the first
point. More precisely, these conditions, which we denote by (A), can be
formulated as follows 
\begin{quote}
\dots given the total state of ourselves and the world at the time along with
the laws of nature, it is open to us to do one thing next and open to us to do
another instead. 
\end{quote}
(Mele (2001), P. 135).

The 'control' mentioned by the second point requires that the choice from the
range of options ought to be 'up to us'. An example of a theoretical idea that
explains more precisely what this means is Mele's (2001, P. 211) 'ultimate
control over $x$-ing by $S$' , where $S$ is an agent and $x$ is an action
(even a mental one): A cause of  $S$'s $x$-ing at $t$ that includes no event
or state internal to $S$ does not exist at any time before $t$. For example,
the causal chain that ends in $x$-ing ought not to start before $S$ was born. 

Determinism and its antithesis are examples of theoretical hypotheses while
condition (A) and the ultimate control are theoretical notions. Theoretical
hypotheses and notions are necessary if we are going to construct a theory
that brings our experience in order and explains it. There is, however, one
rather disturbing feature that the above hypotheses and notions have in
common. 

For example, determinism does not imply any specific observable effects. If we
cannot identify a cause of any event, then there is always the excuse that the
cause exists but is too elusive to be seen. Hence, determinism can be neither
proved nor disproved by any observational data\footnote{Determinism is
  different from the hypothesis that each event from a specific class $C$ is a
  cause of an event from a specific class $E$. In fact, most scientific laws
  have such a form. They can be tested (but cannot be proven).}. 

Similarly conditions (A): How is an experimenter to recognize and describe,
what the total state of the world is? How could he/she reconstruct it at some
other time $t'$ in order to make some results reproducible? This is likely to
be impossible even in principle. 

The hypotheses and notion with such a weak relations to experience are often
called {\em metaphysical}. Moreover, it seems that the above metaphysical
ideas can be identified as the source of the well-known difficulties. An
example (Kane 2005) is the following. Under conditions (A) determinism leaves
only one option so that nobody could have chosen otherwise. The antithesis of
determinism may allow for a non-trivial set of options under the same
conditions but there is no control: nobody can influence random events. 

We shall utilize what may be called the scientific methods (for more detail
see Popper 1972). One idea thereof is the primacy of observational
data. Scientific notions ought to possess sufficiently strong observational
aspects and scientific hypotheses ought to imply sufficiently specific
observable consequences, the so-called predictions, which enable us to test
them. Another requirement is the reproducibility of results so that they may
be tested at different times and by different researchers. Thus, we must keep
metaphysics in reasonable limits. 

In the first part of the paper, we ask whether the science, in particular the
physics, can make do without determinism and what notion could replace
it. Moreover, we would like to see, whether the well-known scientific laws can
admit any kind of freedom. In Sec.\ 2, we find that even Newton mechanics
allows some freedom because it cannot be reduced just to its dynamical
equations. Only after a system and its initial state is chosen, the motion is
determined uniquely by the equations. In quantum mechanics, a similar freedom
of choice is augmented by the statistical character of the theory. This means
that the dynamical laws do not determine the processes uniquely even if a
system and its initial state are both fixed. We shall also see how random
events and choices can be made compatible with (classical) general
relativity. We introduce the important notion of uniqueness of history and
show that space-times of general relativity can be interpreted as histories of
completed time evolutions. 

In Sec.\ 3, the resulting theory of causality is briefly described. The
so-called event causality is shown to be sufficient for all purposes of
physics. The time reversibility of dynamical equations is explained. This
helps to understand the fact that the causality relation of two events is
necessarily oriented and that there is no need to introduce new notions such
as that of state causality. According to Hume, we distinguish the causation
principle from the causality principle. We try to weaken the causation
principle in agreement with the standard interpretation of quantum
mechanics. The structure of the causality principle is found to involve a kind
of conditions, which we shall call relevant conditions. The important feature
of relevant conditions is that their occurrence can be recognized by
observations. With each fixed set of relevant conditions, a definite option
range is associated. Such a range represents a relative freedom, i.e., freedom
with respect to the relevant conditions. Relative freedoms are observable
phenomena and can be encountered in physics, biology and psychology. 

Then we turn to biology in Sec.\ 4 and find that, in a unique and well-defined
sense, living organisms are more flexible than certain class of automata. The
flexibility is based on Popper's concept of trial-and-error processes (see
Popper 1972, Chap. 6, {\it On Clouds and Clocks}). One of the fundamental
features of living beings is the presence of some form of memory. We assume
understanding of memory such as given in the book by Squire \& Kandel
(1999). An important fact about living beings is incomplete knowledge as
represented in their memory. This leads to the existence of the so-called
unforeseen events. Trial-and-error is the method by means of which living
organisms cope with unforeseen events. We describe the ranges of
options---mutations, motions and neural connections---that underlie choice of
trials and we give the account of the corresponding relative freedoms. 

Free will with the three properties listed above seems to be a natural part of
this biological theory. Its only special feature is some conscious
component. Sec.\ 5 conjectures that communicability can be used as a criterion
of consciousness. This has some interesting consequences for the role of
consciousness in free will. Relevant observations and experiments such as
Libet's (2002) are re-interpreted accordingly.

\section{Physics} 
It may be surprising how much the study of physics and its methods will help
us to understand and clarify the notion of freedom.

\subsection{Newton mechanics} 
We start the discussion with Newton mechanics and its structure. To describe
the structure, a few abstract and general notions are needed: the {\em
  system}, the {\em dynamical equation} and the {\em state}. 

The system in Newton mechanics consists of a given number of particles with
given masses, on which given forces act. After a system is chosen the theory
determines its dynamical equation. To obtain a unique motion, a state of the
system must be chosen at a given instant of time, mostly at the beginning of
the motion and then evolved by the dynamical equation. A state is determined
if the coordinates and velocities of all particles are given. 

The choices of a system and a state are not restricted by the theory. These
freedoms are usually understood in a passive sense. For instance, the freedom
of state is seen as the applicability of the dynamical equation to many
different situations in which a system may be found in nature. Similarly for
the freedom in the number of particles, their masses and the forces. It looks
like a paradox but it is just logical: the more general a law is the more
choice it admits. 

Can we go a step further and interpret the choices of a system and a state as
an active freedom of physicists? Such an assumption could be formulated as
follows: {\em Physicists are free to choose a system from a broad class of
  systems and a state of the system from a broad class of the system
  states. Then, they can construct this system in this state in a laboratory
  at an arbitrary time.} However, the words 'physicists are free' have only a
vague meaning here, they are expressing an impression rather than an
established fact. We shall try to make it clearer in Sec.\ 3. 

The impression seems to be always possessed by experimental physicists
(disregard the possible lack of funds). In any case it is not only compatible
with empirical praxis as well as with everyday laboratory work but the
experimental freedom seems to be a tacit assumptions made by scientists
generally. Mathematics and physics are often taught or understood as rigid and
'dead' sets of rules. This is not how these sciences are actually done! 

If we assume that the world in its entirety can be reduced to a system of
massive particles and forces between them so that all properties of all
objects could be calculated from their mechanical parameters according to
Newton mechanics then the future or past state of the world is uniquely
determined by its state at the present time. This is the origin of Laplace's
formulation of determinism. 

According to our contemporary understanding, the assumption that the world can
be described, in all its detail and in its entirety, by Newton mechanics is
wrong. Thus, the theoretical support for determinism provided by it has broken
down. Still, Newton mechanics is a good approximation in many cases and this
makes it an alive and often used theory even today.

\subsection{Quantum mechanics}
Quantum mechanics has the same basic structure as Newton mechanics. Again,
there is an affluence of various quantum systems. With each system a dynamical
equation and a set of states is associated. Given a state at a time instant,
then the state at any other time can be calculated from it by the dynamical
equation and is unique. 

The choice of system and state is not restricted by any rule in quantum
mechanics either. The only difference is that the active version of it is
explicitly formulated in some textbooks of quantum mechanics (see, e.g., Peres
1995, P.50). 

The nature of the states of quantum systems is, however, very different from
that of their Newton counterparts. They must enable a new feature, namely the
{\em statistical character} of quantum mechanics. Given a fixed state of a
system and a quantity that is measurable on the system, then its repeated
measurements generally give different results even if both the state and the
measurement are the same. Only the probability distribution of the results can
be calculated from the state. 

More precisely, a quantum experiment looks as follows. It consists of a number
of individual measurements. In each measurement, we obtain a single quantum
system (for example a photon) from a source (a laser, say), which is some
macroscopic apparatus. The macroscopic structure of the source determines the
quantum state of the system obtained from it uniquely. The (macroscopic)
arrangement of the source and the measuring apparatus (for example a
photographic plate), which is again a macroscopic system, is the same for each
measurement. In each measurement, we obtain a certain value (for example a
position) which can be read off at the measuring apparatus (for example, as a
black point at the plate). If the experiment contains sufficiently many
measurements then the frequencies of the values obtained are well approximated
by the probabilities calculated from the state. 

It is very important to notice that everything done by an experimenter is to
manipulate and observe macroscopic devices. Quantum mechanics only identifies
the cause of the distribution of the measured values: it is the choice of both
a source and a measuring apparatus. But it does not specify any cause for a
particular measurement giving this and not another value. As the values are
visible by naked eye on a macroscopic body, it is the macroscopic behaviour
that is not always predictable.

\subsection{General relativity}
Can the existence of random events and choices be included into a coherent
picture of the whole world? In particular, is it compatible with the rest of
physics? We have seen that Newton mechanics is only valid in a restricted
domain so that its deterministic character does not prevent random events
elsewhere. 

However, there is another modern theory called general relativity. It
describes the world on large scale. An important theoretical concept of
general relativity is that of {\em space-time}: a four dimensional manifold
carrying a Lorentzian metric and containing matter. The structure of general
relativity is similar to that of Newton theory. It admits a number of
different space-times, there are states forming certain sets that are
different for different space-times and space-times must satisfy a dynamical
equation. 

The large scale character of general relativity is important if we are to
compare it with Newton or quantum mechanics. There does not seem to be much
choice of a system if the theory intends to describe the whole
universe. Moreover, we cannot require from physicists to set up an arbitrary
state of the world in their laboratory at an arbitrary instant of
time. However, we can say that different space-times represent different
models of the universe. 

The next problem is that there is no mark on a space-time that would
distinguish the present instant and there is no difference between the general
structures of past and future (this is similar to Newton mechanics). The usual
interpretation is to say that this difference is purely subjective and that
the present instant may be anywhere depending on where observers happens to
be. The space-time is considered as an observer-independent description of the
macroscopic world for all times. 

However, the picture of time that follows from the existence of random events
is dominated by an asymmetry between future and past. The future does not yet
exist and more possibilities are open to it. The past is fixed because the
choice of the possibilities is done only at present instants of time. Such an
asymmetry is not new in philosophy, see, e.g., Popper (1974). 

Consider the past. It can exist only as a memory or another kind of record
that an observer, or a family of observers, can make about the observations
done within each of their progressing presences. Only in this indirect way
does the past have to do with reality. The past as a (processed) record is in
principle fixed in all aspects and details. There are two very different
reasons for that. First, the choice between free alternatives has been done
and no change is possible any more. Second, we usually suppose that different
observations or observations by different observers concerning the choices can
finally be put into an agreement. This is a rather non trivial hypothesis on
which, in fact, all of the science is based. We call this hypothesis {\em
  uniqueness of history}. It has a natural explanation in the philosophical
realism (Russell 1959); for a more contemporary discussion of realism, see
John Searle's (2004) analysis in Chap. 10. 

As for the future, its very existence is a hypothesis based on the analysis
and extrapolation of the past. Similarly, we can extrapolate the existence of
various structures and the validity of laws found in the past. Since there are
also unpredictable aspects, one can say that some part of the world is newly
created (Popper 1974: 'chosen by nature') at each present instant of time,
another  part is determined by the past. 

Now, we can give the answer to the question of how space-times of general
relativity are to be interpreted: Any space-time is just a hypothetical past
of a completed time evolution (Ellis 2006). (A space-time can be viewed as one
possibility for a complete evolution of universe including definite random
choices made at each time). The problem of asymmetry between future and past
then does not arise because we are working only with pasts. If we accept this
change in the interpretation of general relativity, then the existence of
random events becomes compatible with the whole of the contemporary physics.

\section{Causality}
\subsection{Event causality}
Causality and some related notions are very important in discussions about
freedom. In this section, a theory motivated by physics is described. It also
defines our dictionary and tries to prevent confusion caused by the fact that
different texts use the same words in different meanings. 

Let us start by stipulating that causality concerns primarily relations
between {\em events}. An event will be understood as what happens at roughly
one point and at one time (more precisely, in a small space-time
neighbourhood). Thus, an event is not just a small space-time neighbourhood
but something must also happen there. This kind of causality is often called
{\em event causality}. Other kinds of causality can be found in
literature. For example, the structure of a system is considered as the cause
of its properties. It is not clear in how much this has to do with a logical
reason rather than with a physical cause. In any case, such important
relations can be accounted for under the heading of models and theories, not
under causality, and so they will not be lost. Another example of causal
relation that is not a relation between events is the so-called 'agent
causation' (Kane 2005) that occurs in some philosophical texts on free
will. We shall not need it. 

Event causality is sometimes criticized (e.g., Campbell (1957), P.~66) as a
naive, pre-scientific kind of causality and the causality which is met in the
modern field theories is described in a different way: the state of a field at
an instant of time is the cause of another state at a later, or even earlier,
time instant (state causality). Of course, the state of a field at a time is
not an event but we can consider such a state as a set of events\footnote{Even
  in quantum mechanics, any specific preparation of a state can be considered
  as a set of (macroscopic) events.}. Indeed, to describe such a state, the
values of some measurable quantities must be given at all points of space at a
given time, and the fact that these values occur at various space-time points
is nothing but a set of events in our sense. 

The time reversibility of some dynamical equations is also considered as a
difficulty for event causality, which distinguishes  cause from effect and is,
in this sense, oriented. A short reflection however shows that the
reversibility of dynamical equations does not mean the reversibility of the
relation of cause and effect even for state causality. It is true that we can
often reconstruct the state at an earlier moment if we know the state at a
later one. But this is a logical reconstruction. Indeed, we cannot change the
state in the past by manipulating the state at the present. The
re-constructibility may be interesting in that it can be considered as a
criterion of completeness of the event sets which one has to do with. Complete
means here that the set is equivalent to a state, because else the states at
other instants of time would not be uniquely determined. 

Further discussion of the reversibility needs a clarification of the meaning
of this concept first. Indeed, it definitely does not mean that the time flow
can be inverted. At most, the order of {\em some} aspects of a possible motion
can be inverted so that another possible motion results. For example, in
Newton mechanics, the coordinates as functions of time $t$ describe a
trajectory of a motion. One can substitute $-t$ for $t$ into the functions and
another possible trajectory will result (if the forces are time-reversal
invariant). Now, the states of Newton mechanics are completely determined only
if both coordinates and velocities are given. But the velocities at the
corresponding points after time reversal are not the same as before, they must
also be inverted. Hence, the same states are not running in the reverse order
if we 'reverse time' and the final state of the original motion cannot be the
cause of its initial state. To summarize, it seems that some version of event
causality is an adequate notion, and it is even more satisfactory for physics
than various reversible state causalities. Let us elaborate on the
corresponding causal relation in more detail. 

A {\em causal relation} is an oriented relation between a set of events
$\mathcal A$ called the {\em cause}, and an event $B$, called the {\em
  effect}. If $\mathcal A$ occurs, $B$ must also do ($\mathcal A$ is called by
some philosophers 'sufficient cause'). For example, in Newton mechanics, the
elements of a cause $\mathcal A$ can be values $x_n^i$ of the three
coordinates and velocities $\dot{x}_n^i$, $i=1,2,3$, for all particles
$n=1,2,\cdots,N$ at some time $t$. Then, $B$ can be certain pair ${x_n^i,
  \dot{x}_n^i }$ concerning $n$-th particle and $i$-th coordinate axis at a
time $t'$ larger than $t$. If we take all such $B$'s at the time $t'$, then we
have a complete system of effects so that we can reconstruct $\mathcal
A$. This suggests that we can collect several effects of the same cause in a
set and call this set also an effect of the cause. But, unlike causes, the
composed effects seem to be always decomposable into their event elements. 

{\em Locality} means in physics that causes always precede their effects in
time, $t({\mathcal A}) < t(B)$. In modern physics, this mostly implies that
the effect must lie in the future light cone of the cause: causing does not
travel with a velocity larger than the speed of the light. There are some
exceptions, but these are always due to some special circumstances that can be
detected. For example, in the case of  super-luminal group velocities in some
media etc., see Liberati, Sonego \& Visser (2002), there always seems to be an
inertial frame where no effects precede their causes, such as the rest frame
of the medium.

\subsection{Principle of causality}
Thinking about causality was strongly influenced by Hume (1992). Hume
distinguished two hypotheses: the {\em principle of causality}, which states
that like causes have like effects, and the {\em principle of causation},
which maintains that there is no event without cause. Hume also listed all
properties of the causal relation given above. 

The principle of causality seems to be the most important part of the theory
because without it the observation of causal phenomena would be
impossible. Let us try to give the principle a more precise form. First, we
say 'equivalent' instead of 'like': equivalent causes and equivalent effects
form classes that are determined by some conditions. 

Physics and its ways can help to understand the conditions. First, the objects
which physicists work with are called systems. Systems usually come in many
copies and there is some way to recognize that two systems are equivalent. 

Second, physicists study systems under specific conditions that must be
reproducible (in a laboratory, say) or at least recognizable and often
encountered (in astrophysics, say). The conditions should not in general
contain the time and the position. This enables us to observe equivalent
causes at different times and positions and to check that there is a
pattern. That is, observations can be repeated in order to confirm or extend
previous results. We also assume that every occurrence of such conditions
consists of a set of events. The conditions do not have any special name in
physics, but for our purposes it is advantageous to give them one. Let this be
{\em relevant conditions} because they are indeed relevant for the study or
the experiment. 

In fact, the notion of relevant conditions is the basic tool of our
understanding the world around us. All laws of nature have been discovered by
looking what happens if some relevant conditions occur repeatedly. In this
way, each particular causal relation becomes an instantiation of a general law
of nature\footnote{Causal relations and laws of nature cannot be proved by
  pure logic from pure observational evidence. This was also shown by Hume. We
  can agree with Hume by saying that they remain only more or less probable
  hypotheses, but we still maintain that this does not make them useless.}. 

The science makes a heavy use of relevant conditions for example in its
experimental methods. The experiments are to be reproducible and so their
conditions have to be defined carefully. It seems, however, that the
recognizability and reproducibility of the conditions is never absolute; there
is no absolutely clean experiment. There always seem to be uncontrolled
influences; in any case, repeated experiments always give some dispersion of
numerical values. The lack of an absolute control about the purity of the
conditions can be quantified by the dispersion (mean squared deviation). In
practice, one views a control as excellent if the dispersion is sufficiently
small.

\subsection{Principle of causation}
Let us turn to the principle of causation. Historically, this is a very old
principle. For example, almost the same words were used by Hume and Platon
(see the Introduction). From the principle of causation, it follows that we
could predict everything that happens in the future of a time instant, if we
knew all kinds of causes and if we knew everything that happened in the past
of the time---determinism. Determinism can also be described as follows: Given
the complete state of the whole universe at a time instant $t$, then there is
only one possible way of how the universe evolves in the future or past of
$t$. 

It has been explained in the Introduction that the principle of causation and
the associated determinism cannot be proved or disproved. What we can control
in physical experiments with, and in observations of, a system is always only
a tiny part of the universe so that we can recognize if it satisfies some
relevant conditions. A system subjected to given relevant conditions exhibits
a behaviour. If the conditions are repeated, the system can show the same or a
different behaviour. In this way, one can find a number of alternative
possibilities that are open to the system under the given conditions. If,
after many repetitions, no new possibilities occur we can assume, that they
have all been found. Such a complete system of possibilities is called {\em
  option range}. The existence of the options can, therefore, be
experimentally tested. The knowledge about specific option ranges form an
important part of science. 

Most contemporary physicists simply assume what we call the {\em Weak
  Causation Principle} (WCP) and what we can formulate as follows: {\em Every
  event is an element of an option range determined by some maximal relevant
  conditions}. 

Relevant conditions are called {\em maximal} if adding any further condition
does not decrease the number of options. If the corresponding option range
contains only one element we have a causal relation of the usual type. But WCP
admits also cases when the range contains more alternative possibilities. The
existence of such cases is assumed by the standard interpretation of quantum
mechanics. WCP does not comprise much more than just what experience tells us
but it is sufficiently strong to explain our search for causes, interesting
relevant conditions and option ranges. 

After about 80 years of quantum mechanics, WCP seems to be a simpler and more
comfortable hypothesis than the determinism. To introduce additional causes
(in quantum mechanics, the so-called hidden variables) into the theory makes
it technically more complicated. Moreover, in this way, something is
introduced that does not have even an indirect relation to what can be
observed today. By Occam's razor, we are better to abstain. If evidence will
emerge sometimes in the future that will have any relevance to the problem, it
will most likely be very different from anything what we could imagine in our
wildest dreams today. However, it is not true that quantum mechanics has
proven determinism to be wrong.

\subsection{The notion of relative freedom}
The uniqueness of history makes determinism impossible to disprove and freedom
difficult to see. Still, physical experiments show the way of how a rational
notion of option ranges can be introduced. Crucial is the concept of relevant
conditions. They are weaker than conditions (A) mentioned in the Introduction
and posses a closer relation to observational data. 

We can define: {\em If a system has an option range that contains more than
  one element under some fixed relevant conditions, then we say that it has a
  freedom relative to the conditions, a relative freedom}. Hence, one and the
same system can have several different relative freedoms depending on the
choice of the relevant conditions. 

Our definition of option ranges does not mention the way in which its
alternative possibilities are realized. It can be an effect of a cause that
lies outside the relevant conditions (then the conditions are not maximal) or
it need not have any cause. The way of the realization of a fixed option may
even differ in different cases of its realization. The existence of relative
freedoms itself contradicts neither the existence of causes nor that of random
events. 

Let us stress that relevant conditions may define an important and useful
relative freedom even if they are not maximal. An example is connected with
the so-called emergent phenomena. These are properties of complex systems that
cannot be derived exclusively from the properties of its individual
constituents. The possibilities of how given constituents can combine form an
option range. Consider for instance electrons, protons and neutrons. They can
combine into about one hundred stable atoms that in turn can combine into
zillions of stable molecules, crystals and mixtures. This is a huge option
range, which underlies the surprising wealth of structure in nature. 

A noteworthy example of non-maximal relevant conditions is the relative
freedom that is associated with each scientific law such as dynamical
equations in physics. Indeed, the more general a law is, the larger the number
of individual cases can be ruled by it. If we define the relevant conditions
as the applicability of the law, then the individual cases form the option
range of a relative freedom. The ability to use the law entails a deep
knowledge of this freedom. 

We now have a better understanding of the impression about the freedom of
physicists referred to in the subsections on Newton and quantum
mechanics. There are relevant conditions: to have a well-equipped physical
laboratory, full stop. There is an option range: all mechanical or quantum
mechanical systems that can be constructed in the laboratory and all states,
in which the systems can be created. First, it is clear that the statement can
be given an arbitrarily rigorous form, second, it expresses exactly the
freedom that is necessary to do physics and, finally, it seems that it is an
adequate explanation of the freedom impression of the physicists.

\section{Biology}
Let us define the living organisms provisionally by counting, just taking over
the existing listing and taxonomy. Experience shows that living beings all
have what we can call {\em elementary needs}: they feed, they reproduce and
they avert or avoid danger (we understand these words in a sufficiently
general sense so that they also apply to all living organisms). Then, their
behaviour can be understood as ultimately motivated by the elementary
needs. There is therefore what can be called {\em teleonomy}. The teleonomy is
'built in the genes'. An account of it is given by Monod (1970). 

Of course, there are organisms that have much more sophisticated
motivations. It seems, however, that the 'higher' motivations can be
understood as {\em derived needs}, needs that have their origin in strategies
serving to satisfy the elementary needs. For example, some animal species are
gregarious, and it is more or less clear, how a suitable way of life in a
group makes feeding, reproduction and averting dangers easier or even possible
in a given niche. Then, seeking, gaining and keeping a suitable position in a
group may be understood as a derived need. 

Another important property of the known living organisms is that they are more
flexible than the machines that follow fixed algorithms. A generally used
mathematical model of such a machine is the so-called Turing (1937)
machine. From the biological point of view we can describe an algorithm as
follows: it determines specific responses to specific stimuli and there is a
fixed list of the stimuli (a list of the responses is determined by the
organism)\footnote{Even (artificial) neural networks (see e.g. Churchland
  (1995)) that are capable of learning can be simulated on digital computers,
  i.e., in principle on Turing machines. The reason is that they can only
  learn to associate a response from a fixed closed list with a stimulus from
  another fixed closed list.}. In our language of conditions and options, we
can say that there are conditions---identified with always one specific
stimulus of the list---that possess only one option, a response from the other
list. To be sure, the behaviour of all living organisms including people
follows in some stages some such program. Let us call it {\em
  routine}. However, the strategies of living species are not exhausted by
routines. On the other hand, we do not claim that it is impossible to
construct a machine that is as flexible as a living organism. Observe that the
lists of stimuli and reactions represent information, e.g., some knowledge
about how the stimuli are related to the elementary needs. 

Suppose that an event occurs that is not in the list of stimuli, but it none
the less still has relation to the elementary needs. Let us call such an event
{\em unforeseen}. We leave open whether an unforeseen event has had a cause or
not. The existence of unforeseen events can be interpreted as an instance of
{\em incomplete knowledge}: the knowledge is missing that concerns both the
link of the unforeseen event to the elementary needs and the way in which this
link can be used for a suitable reaction. The memory of an organism does not
contain any direct representation of such knowledge. 

The flexibility means that all species are equipped with facilities (body
structures) that enable them to reprogram the routines if some unforeseen
events occur. An important property of unforeseen events is that there does
not exist any list of them. Hence, a solution to such a problem cannot be
programmed in the way Turing machines work. A general procedure serving this
purpose that can be observed in all living beings has been highlighted by
Popper (1972), who has called it the '{\em trial-and-error} method'. The same
principles have been applied by Edelman (1987) just to the way brains are
working. However, his 'neural Darwinism' is aimed at perceptual
categorization, generalization and memory rather than at free will.  

In our language of conditions and options, we can give the following  account
of the 'method'. Consider an organism such as a mouse and let some specific
conditions contain an unforeseen event and let the conditions allow some range
of options for the mouse response. Then, the mouse that is confronted with the
conditions must be able to make choices from the range. Let us call the
procedure of such choice {\em realization}. The realization process may be
random or causal or something in between---this is irrelevant for our
theory. The ability to perform the realization process needs freedom that is
possessed by all living beings. Pointedly said, it is the liberty to
err. After each choice, the mouse perceives what happens. Often, the result is
negative (error). The next time when the conditions occur, the mouse makes
another choice. And so on, until the result is positive. The good choice may
then be remembered and the routine thus reprogrammed. This part of the
procedure is called {\em selection}. 

It is clear that a necessary condition for the procedure of trial and error to
work is some kind of {\em memory}. The errors have to be remembered so that
they will not be repeated. The selections have to be remembered, in order that
they will be chosen. And indeed, some kind of memory is the most important
part of any living being.  

Our theory of biological freedom explains the frequent observation that
structures evolved to serve certain needs are subsequently used for very
different puposes. Let us now look to see how it works in more detail.

\subsection{Mutations}
The most basic kind of memory in biology is an inheritance molecule such as
the deoxyribonucleic acid (DNA) and the basic option range in biology is
formed by {\em mutations}. Its alternative possibilities are the changes in
the DNA. Mutations come about in a realization process that may work all the
time (as, e.g., for bacteria) or during specific periods (meiosis for sexually
reproducing organisms; a generally understandable description thereof is
Dawkins 1989) and it seems to be random. 

If the option range is wide enough then some of its possibilities are
advantageous for the mutated organism in the sense that they lead to the
proliferation of individuals with this mutations and to relative suppression
of the others. In this way, the selection process works. As a rule, the
selection is not random and causes can be found for most selections. 

Observe that the unforeseen event can be answered only if it comes  after the
suitable mutation is established and remembered. For example, if a new
antibiotic is applied to a bacterial infection, there must already be bacteria
with a suitable mutation. Thus, in the special case of mutations, the
unforeseen events themselves take part in the selection process.

\subsection{Neuron connections}
Some multicellular organisms such as animals possess an option range with
alternative possibilities called {\em motions}. Parts of the animal's body,
e.g., trunks, legs or eyes, can take different relative positions without
inhibiting other functions of the body. The relevant conditions are simply the
internal anatomy and physiology of the body as well as the external
circumstances that allow changes of such positions. 

Animal motions are usually organized with the help of the nervous
system. Properties of nervous systems are important for us but a simplified
picture that follows will be sufficient. Much more detailed neurology
supporting our approach can be found in Edelman's (1987) 'theory of neural
groups selection.' 

Experiments show that certain nerve signals trigger certain motions. Sequences
of motions such as running or flying are brought about by specialized sets of
nerves connected in a particular way. Influence of sensory data on motions is
made possible by connections of other sets of nerves. The connections of
neurons can thus code for processes containing motions and can themselves be
considered as a memory. Some neural structures are inherited: the connections
are built up according to the DNA blueprint. This kind of memory serves for
routines and cannot be used for trial-and-error processes. The connections
have been selected by the process based on the mutations and the contents of
the neural memory is equivalent to a part of the DNA. 

Other connections of neurons can, however, be altered also during the life of
an individual organism. Let us consider changes enabled by the so-called
synapses. The alternative possibilities of how the strengths of synapses can
be chosen form an important option range for animals. The relevant conditions
here are the positions and variation range of all flexible synapses. The
simplest example of such an option range and a memory based on it can be
recognized in the learning by invertebrates as described by Squire \& Kandel
(1999). There are three kinds of such learning:  habituation, sensitization
and classical conditioning. Let us look at the habituation. The other two
kinds of simple learning are of similar nature. 

Let some new stimulus that does not carry any danger by itself cause some kind
of alarm response by an animal. In this case, we can arrange a training
session in which this stimulus is occurring more often without being followed
by anything harmful. Then, the response becomes gradually weaker until it
practically disappears. This is first remembered only for a short time of,
say, ten minutes. If the training sessions are repeated sufficiently many
times over a sufficiently long time, then the new reaction can be remembered
for a longer time, such as several weeks. 

The connectivity of the neurons as well as the physical and chemical processes
in the synapses that are necessary for such learning are well understood. We
can describe the learning in our language in agreement with these neurological
details. The relevant conditions are the occurrence of the stimulus and no
harmful event following it (the unforeseen event). The options are the degrees
of alarm. The realization process is carried out by the chemistry of the
synapses and is rather systematic: the degree of alarm is approximately
decreasing. Observe the important role of the repeated occurrence of relevant
conditions so that an animal can perceive what its trials are leading to. The
no-alarm option is selected after a chain of sensory inputs. What has been
learnt is a specific response to a specific stimulus. This is not just a kind
of 'second-order' Turing machine: true, a response to a stimulus chain is a
change of a response to a stimulus. But, as it was already stressed, there is
no list of unforeseen events. The method is trial-and-error. 

More advanced kind of learning and trial-and-error processes are enabled by
the so-called declarative memory. It needs a collaboration of two complex
nervous structures: the cortex and the hippocampus. Consider for example the
experiments with rats in the Morris water maze (Squire \& Kandel 1999). This
is a large circular pool with murky water and steep walls that cannot be
climbed by the animal. Under the water not at the centre of the pool, a
slightly submerged platform is hidden. 

The rats are put into the water at some fixed position in the pool. The
subsequent chaotic motion of a rat in searching for something (it does not
know about the platform) has the character of a trial-and-error process. The
options are nervous signals that code for and trigger different possible
swimming motions. We do not need to know in detail the nature of these signals
and simply call them {\em portable neural representations} (PNR). They seem to
occur in a rat's brain in a more or less random way. Some of them may be
related to some stimulus from outside as well as from the rat's memory. The
realized PNR's that lead to motions are the trials of the process. Finally,
the successful PNR representing a more or less direct swim to the platform is
selected and remembered. 

After this training, the experiment is modified in order to see what kind of
knowledge has been acquired about the location of the platform. There are two
groups of rats: those with healthy hippocampus and those with lesions
there. The rats of both groups are able to swim directly to the platform from
the position where they have been always put into the water during the
training. Next, the rats have to start from another position in the pool. It
turns out that the normal rats swim to the platform from any position without
searching while those with lesions have to fall back on trial and error. 

To explain these observations, one possible assumption is that the healthy
rats acquire more knowledge about the layout of the room than the others. The
knowledge is in the form of a PNR that somehow represents the layout. It is
possible in principle because the room with the pool is not rotationally
symmetric: its walls carry some cues. The layout PNR can be used by the rats
to work out the suitable motion from the sensory data on where they have been
put into the water. If this calculation takes the form of a purely mental
trial-and-error process (see Edelman 1987), then the new starting position
{\em is} an unforeseen event and what the healthy rat does is {\em not} a
routine. If not, then the healthy rats have simply learned a better kind of
routine than the others. 

Another hypothesis is that the acquired layout PNR is similar for both groups
but the rats with lesions cannot work out from it the correct motion. There
are two possibilities. First, the purely mental trial-and-error process cannot
be performed without hippocampus. Second, the routine calculation cannot be
done so. In this case, the lesser calculation ability has the same effect as
the incomplete knowledge. 

By the way, we observe that none of the rats chooses to swim merrily around
instead of searching for something. They apparently have an inbuilt dislike
for staying in water that has some connection to their elementary needs.

\section{Freedom of will}
The freedom of will is usually understood as the ability to choose consciously
an idea of an action and then carry out the action accordingly. The general
notion also agrees with free will having the three properties listed by Double
(see the Introduction). In this section, we try to construct a theory
explaining these points. 

Clearly, the choice is only possible if there is an option range corresponding
to some conditions. Our main idea is to replace conditions (A) by the relevant
conditions. In this way, our theory of free will becomes a part of the general
biological theory described in the previous section. The only new feature that
makes a relative freedom to free will is a suitable conscious component. 

As for ideas, we assume that each is associated with a portable neural
representative\footnote{The ideas are mental states or processes and PNR are
  neural states and processes. We accept the solution of the mind body problem
  as described in detail by John Searle (2004). Roughly, the ideas and PNR are
  different entities, which, however, can be considered as two sides of the
  same thing. Similar thoughts can be found already in Fechner (1860) and
  perhaps even before Fechner.}. 

The nature and role of consciousness and how the consciousness is represented
by neural processes in the brain is not known. However, some incomplete and
phenomenological understanding might be sufficient for our purposes. In
particular, an important criterion of consciousness is the communicability: a
conscious idea, or that an idea is conscious, can be told to others. In fact,
this criterion is crucial for the experimental work done e.g.\ by Libet: there
is no way experimentalists can be informed on consciousness of another person
independent of what the person tells them. 

There are various phenomenological hypotheses on consciousness. Some
surprising observational data (see Wegner 2002) suggest (but do not prove) the
conjecture that consciousness is only an epiphenomenon. If this were true,
then the free will in our sense would still exist: we could define it as
conscious {\em experience} of relative freedoms. However, it seems that the
data admit a stronger conjecture. 

If we adopt the communicability criterion, then it follows that consciousness
helps us actively to carry out complex mental work. A necessary component of
such a work is a formulation of partial results from time to time in a way
that is understandable to others even if no communication is planned. This
must be conscious according to the criterion. The purpose is to check that
nothing has been forgotten and that the whole is logically coherent. We can
thus try the more general conjecture that consciousness provides active and
manifold help with thinking. It is also supported by the observation that
consciousness is always experienced in non trivial processes of
thinking. Moreover, it explains why brains with consciousness have advantage
over those without and why they, therefore, would evolve at all. 

As free will is concerned, the conscious component in realizing and selecting
ideas carries with it both advantages and disadvantages. On the one hand it
helps thinking, on the other, it consumes more energy and time than
unconscious processes. Indeed, the relatively long time needed for conscious
processes is an established observational result (Libet 2004, Wegner
2002). Brain processes consume glucose at a pretty high rate. This suggests
that there is also need for more energy. If conscious processes in the brain
run parallel to unconscious ones or if they entail synchronous actions of many
neurons (Llin\'{a}s \& Ribary 1993), then they are expensive. 

Our experience forces us to admit that unconscious processes take also part in
thinking. What is their role? Let us look more closely at the experiments
supporting the conjecture that actions are chosen unconsciously and that
consciousness is only an accompanying phenomenon. A conspicuous feature of the
experiments is that the actions asked for do not require any complex mental
work. For example Libet asks for the choice of an arbitrary time instant for a
simple motion of a wrist. 

Hence, a simple alternative interpretation of such experiments is as
follows. When the experiment is running, the unconscious brain does not switch
on the consciousness at once because no complicated mental work is needed. It
does the choice itself, only then 'drops a notice' to the consciousness to
enable a possible veto and to the declarative memory to save it for later,
maybe conscious, use. 

Let us instead look at the game of chess. This is an activity that clearly
needs non trivial thinking. First, a chess player has motives and desires to
play or not to play a game, to win or not to win as well as some more special
tastes of how to play. Some vague ideas about connections of these thoughts to
the elementary needs could be imagined, but we leave the question open. In any
case, we do not require the origin of the desires to be always a conscious,
rational reasoning. 

After the decision to play and win and after the start of a game, there are
well defined option ranges: all moves that are allowed by the rules of chess
in a given position. The moves are understood as ideas. Some positions with
which a player is confronted during the game are unforeseen events as defined
at the beginning of Sec.\ 4. 

No chess player is able to calculate the game to the end from most positions;
not even the fastest computer can do this in a reasonable time. The players
are looking for a 'strong' move that is enabled by some properties of the
position. If they become aware of any such move they calculate the possible
consequences of it to the depth of several moves to check if the idea
works. In this way, they learn more about the position so that after rejecting
one idea, they are likely to get another, etc. It is not allowed to make trial
moves with the pieces, hence we have a purely mental method of trial and
error. Conscious processing is influenced by input coming from the unconscious
brain. The players become aware of relations to other games that they played
or studied in the past. The process of selection from trial moves is an
example of process in which the help of consciousness can be experienced at
many intermediate steps. 

The final point is that after having calculated the moves and selected among
them with the help of consciousness the act of actually moving a piece may
again contain unconscious elements. Indeed, it is often done in an utterly
mechanical way that is not even remembered. The question whether the conscious
mind directly moves the hand seems to be less important than whether the hand
moves in accordance with the mind plan. 

The chess example makes an important point. Consciousness is considered as a
tool helping to choose from option ranges in complicated cases. The manner of
its working as such a tool explains the third point by Double (see the
Introduction): some rational and sensible aspects of free will. Indeed, as we
have argued, one of the tasks of consciousness is to check that the selection
is based on a logically coherent system of ideas and that it satisfies the
desires. 

A still stronger conjecture on the role of consciousness is a hypothesis that
I would call the {\em consciousness domination}. The idea is that, on the one
hand, desires and motivations are formed rationally in a conscious way and, on
the other, that conscious thoughts can be direct causes for actions. Wegner
(2002) had thoroughly discussed the second part of this hypothesis, which he
called the {\em conscious will}. He had drawn on a wealth of experimental and
observational data and concluded that this kind of conscious will is an
illusion. Now, the analysis of the chess game shows that the theory of free
will explained in the present paper does not make use of any part of the
consciousness-domination hypothesis. 

Let us turn to the second point by Double. A scientific theory of control can
go roughly along the following lines. It starts at the observation that all
living organisms are born with some teleology: there are certain elementary
needs that have to be satisfied. Humans are gregarious animals, with an
involved hierarchical social structure. They need to find, gain and keep a
suitable position in society as was mentioned at the beginning of Sec.\ 4. The
needs are clearly given to us from nature and are not up to us. But the
inbuilt teleology shows only general directions. In fact, we are overwhelmed
by the huge manifold of choices to which different people are stimulated by
their inherited teleology. Let us call the basic choices of this kind the {\em
  options of self}. 

On the one hand, the choices from the options of self are not completely up to
us. Clearly, some of them develop even without our conscious acting through
breeding, education, accidental circumstances, etc. On the other hand, the
corresponding desires and motivations need not even be compatible with each
other and one has to deliberate and choose. Moreover, many of us have the
experience of, as it were, consciously and freely changing a quite essential
part of the self after some very important unforeseen event happens, although
such events are rather rare. The choice must be done in agreement with our
feeling of what is the best for us and this is the basis of some
control. Next, a more or less complete, definite and constant self does not by
itself determine uniquely our specific everyday choices. Again, there can be
conflicting desires on the one and different ways of satisfying one desire on
the other hand. The control means that we make our choices in the best
possible accord with our self if an unforeseen event occurs. 

If these ideas on control are properly interpreted then they can be recognized
as a part of the teaching on the self in the contemporary psychology, see
e.g.\ Baumeister (1998). The difference between the present paper and the work
of psychologists is that the research in psychology focuses on patterns of
experience and studies the behaviour that is a consequence of our impression
of being free to choose. Hence, a lot of useful work can be done without
addressing the philosophical problem associated with a {\em real} free
choice. The result is a purely phenomenological theory that does not explain
its basic notions. Some do just that (Baumeister), others interpret the
impression of free choices as an illusion (Wegner). 

Our biological definition of free will makes it a phenomenon that is
compatible with all scientific evidence and can be studied by natural sciences
on the one hand and that seems to agree with everyday impression of the
freedom as we all know it on the other. However, only some rough general idea
has been be given here and many details still remain to be elaborated on.

\subsection*{Acknowledgements}
The author is indebted to Uve-Jens Wiese, Gilberto Collangelo, Victor Gorge,
Claus Kiefer, Karel Kuchar and Oldrich Semerak for reading some of earlier
version of the manuscript and for suggesting improvements. Thanks go to the
Institute of Theoretical Physics, Faculty of Mathematics and Physics of the
Charles University, Prague for hospitality and discussion.

\section*{References}
Baumeister, R. F. The Self. In {\it Handbook of Social Psychology}, 4th
edition, eds. D. T. Gilbert, S. T. Fiske and G. Lindzey. Vol. 1,
680—740. Boston: McGraw-Hill.\newline 
Churchland, P. M. {\it The Engine of Reason, the Seat of the Soul}. Cambridge,
Mass.: The MIT Press.\newline  
Campbell, N. R. (1957) {\it Foundations of Science. The Philosophy of Theory
  and Experiment}. New York: Dover Publications. \newline Dawkins, R. (1989){
  \it The Selfish Gene}. New York: Oxford University Press. \newline  
Double, R. (1991) {\it The Non-Reality of Free Will}. New York: Oxford
University Press. \newline  
Edelman, G. M. (1987) {\it Neural Darwinism. The Theory of Neuronal Group
  Selection}. New York: BasicBooks. \newline 
Fechner, (1860) {\it Elements of Psychophysics}. New York: Holt, Rinehart and
Winston, 1966. \newline 
Ellis, G. F. R. (2006) Physics in the Real Universe: Time and
Spacetime. Arxiv, gr-qc/0605049. \newline  
Hume, D. (1992) {\it Treatise of Human Nature}. Amherst: Prometheus
Books. Book I, Part III. \newline  
Kane, R. (2005) {\it A Contemporary Introduction to Free Will}. New York:
Oxford University Press. \newline  
Laplace, P. (1820) {\it Th\'{e}orie analytique des probabilit\'{e}s}. Paris:
Courcier Imprimeurs. \newline  
Liberati, S., Sonego, S., \& Visser, M. (2002) Faster-than-c Signals, Special
Relativity and Causality. Ann. of Phys. {\it 298}, 167--185. \newline  
Libet, B. (2004) {\it Mind Time. The Temporal Factor of
  Consciousness}. Cambridge (Massachusetts): Harvard University
Press. \newline  
Llin\'{a}s, R. and Ribary, U. (1993) Proc.\ Nat.\ Acad. Sci.\ (USA) {\bf 90}
356. \newline 
Mele, A. R. (2001) {\it Autonomous Agents. From Self-Control to Autonomy}. New
York: Oxford University Press. \newline  
Monod, J. (1970) {\it Le hasard et la n\'{e}cessit\'{e}}. Paris: Edition  du
Seuil. \newline  
Peres, A. (1995) {\it Quantum Theory: Concepts and Methods}. Dordrecht:
Kluver. \newline  
Popper, K. R. (1972) {\it Objective Knowledge}. Oxford: Clarendon
Press. \newline  
Popper, K. R. (1974) {\it Unended Quest. An Intellectual Biography}. London:
Fontana and Collins. Chapt. 28, Meeting Albert Einstein. \newline  
Russell, B. (1959) {\it The Problems of Philosophy}, Oxford: Oxford University
Press. \newline  
Searle, J. R. (1984) {\it Minds, Brains and Science}. Cambridge
(Massachusetts): Harvard University Press. \newline  
Searle, J. R. (2004) {\it Mind. A Brief Introduction}. New York: Oxford
University Press. \newline  
Squire, L. R., \& Kandel, E. R. (1999) {\it Memory. From Mind to
  Molecules}. New York: Scientific American Library. \newline  
Turing, A. M. (1937) Proc.\ Lond.\ Math.\ Soc.\ (ser.\ 2) {\bf 42} 230;
correction {\bf 43} 544. \newline 
Wegner, D. M. (2002) {\it The Illusion of Conscious Will}. Cambridge, Mass.:
The MIT Press.\newline  
\end{document}